\documentclass[twocolumn,twocolappendix]{aastex631}
\usepackage{amsmath}

\usepackage{CJK}

\shorttitle{galaxy pair in filament}
\shortauthors{Peng Wang}

\begin{document}
\begin{CJK*}{UTF8}{gbsn}

\correspondingauthor{Peng Wang}
\email{pwang@shao.ac.cn}

\title{Beyond the Local Group - I: Cosmic Filaments Govern the Spatial Alignments of Galaxy Pairs}

\author[0000-0003-2504-3835]{Peng Wang* (王鹏)}
\affil{Shanghai Astronomical Observatory, Nandan Road 80, Shanghai 200030, China}

\begin{abstract}

Using a large observational sample from the Sloan Digital Sky Survey, we investigate the spatial alignment between galaxy pairs and their local cosmic filaments. Focusing on pairs with stellar masses and separations comparable to the Milky Way-Andromeda (MW-M31) system, we measure the angle between the pair connecting line and the orientation of the host filament, determined using a filament catalog constructed via the Bisous model. Our analysis reveals a statistically significant tendency of galaxy pairs to align their connecting lines along their host filaments, manifesting as an overall $\sim$7\% excess of alignment angles smaller than the MW-M31 case compared to a random distribution. Crucially, the strength of this alignment exhibits a strong dependence on the distance to the filament spine. Pairs located within 0.2 Mpc from the filament spine show the strongest alignment, while those beyond 1 Mpc display no significant alignment. Furthermore, we identify a bimodal distribution of alignment angles near filament cores, suggesting distinct dynamical populations potentially associated with infall and interaction processes. Our results provide robust observational support for theoretical models in which anisotropic accretion and tidal forces within the cosmic web drive galaxy pair evolution. They also position the MW-M31 system as a representative filament-aligned pair, offering insights into Local Group assembly. This study demonstrates the cosmic web's critical role in dictating pair orientations and motivates future work with kinematic data to unravel galaxy-environment interplay.

\end{abstract}

\keywords{
    \href{http://astrothesaurus.org/uat/929}{Local Group (929)};
    \href{http://astrothesaurus.org/uat/902}{Large-scale structure of the universe (902)};
    \href{http://astrothesaurus.org/uat/330}{Cosmic web (330)};
    \href{http://astrothesaurus.org/uat/2029}{Galaxy environments (2029)};
    \href{http://astrothesaurus.org/uat/595}{Galaxy formation (595)}
}

\section{Introduction} 
\label{sec:intro}

The matter distribution on large-scale of the Universe, often visualized as the cosmic web, is composed of a vast network of clusters, filaments, sheets, and voids \citep[e.g.,][and references therein]{1986ApJ...302L...1D,1970A&A.....5...84Z,1978IAUS...79..241J,1996Natur.380..603B,2004MNRAS.350..517S,2013MNRAS.429.1286C,2014MNRAS.441.2923C,2018MNRAS.473.1195L}. Dark matter halos and the galaxies they contain are not randomly distributed in the cosmic web \citep[][]{2005Natur.435..629S, 2010MNRAS.408.2163A}. Instead, they tend to cluster along filaments and within dense nodes \citep[][]{1983MNRAS.204..891K,2006Natur.440.1137S}, leading to a diverse range of environments that profoundly influence galaxy evolution. Recent progress in wide-field spectroscopic surveys \citep[e.g.,][]{2000AJ....120.1579Y,2016MNRAS.460.1270D} and high-resolution numerical simulations \citep[e.g.,][]{2014Natur.509..177V} has enabled astrophysicists to investigate extensively the cosmic large-scale structure and its influence on galaxy properties.

It is well established that a galaxy's morphology \citep[e.g.,][]{1980ApJ...236..351D,2004MNRAS.353..713K}, size \citep{2019RAA....19....6Z}, star formation rate \citep[e.g.,][]{2010ApJ...721..193P,2009ARA&A..47..159B}, and shape or spin axis orientation \citep[e.g.,][]{2012MNRAS.427.3320C,2013MNRAS.428.1827T,2013ApJ...762...72T,2014MNRAS.443.1090F,2017MNRAS.468L.123W,2018MNRAS.473.1562W,2024MNRAS.528.4139H,2025ApJ...983..122R,2025ApJ...983L...3R} are closely linked to their location within the cosmic web. Moreover, intrinsic alignments of single galaxies' spin axes with the cosmic web have been observed across a wide range of spatial scales and persist up to high redshifts \citep[e.g.,][]{2015MNRAS.448.3391C,2014MNRAS.444.1453D,2020MNRAS.491.4294K,2024PhRvD.109l3548S}. This environmental influence is not confined to individual galaxies but also extends to small galaxy systems such as pairs \citep{2015A&A...576L...5T}, and triplets \citep{2024MNRAS.531L...9R}, suggesting that the cosmic web shapes both the internal properties of galaxies and their spatial configurations within groups. These small systems exhibit intrinsic alignments with the local tidal field traced by the cosmic web, highlighting the significant role of cosmic environment in organizing galaxy systems beyond individual scales.

A particularly intriguing example is the Local Group (LG), composed of the Milky Way (MW), Andromeda (M31), and their satellite systems. Several studies have shown that the MW-M31 pair is embedded within a local filamentary structure \citep[e.g.,][]{2015MNRAS.452.1052L,2015MNRAS.450.2727T}, and the spatial alignment between the MW-M31 connecting line and the orientation of this filament may reflect anisotropic accretion and the influence of the cosmic web on the assembly of the Local Group \citep{2022MNRAS.516.4576D}. More broadly, \citet{2015A&A...576L...5T} provided direct observational evidence for a statistically significant alignment between galaxy pairs and their host filaments using Sloan Digital Sky Survey (SDSS) data \citep{2000AJ....120.1579Y,2014ApJS..211...17A}. In their analysis, galaxy pairs were selected based on projected proximity and velocity criteria, and only those pairs whose center points lie within filaments identified by the Bisous model \citep{2016A&C....16...17T} were considered. The angle between the pair connection and the local filament orientation was measured in the plane of the sky to minimize redshift-space distortions. Their results revealed a clear excess of galaxy pairs aligned with filaments compared to a random distribution, with the alignment signal being especially strong for ``loose pairs''—those with larger separations.

While some progress has been achieved, key questions persist regarding the physical origins of these alignments and how they vary with galaxy properties and environmental factors. This issue is particularly pertinent for systems like the LG, whose unique configuration provides a critical testbed for theories of galaxy formation and the influence of the cosmic web. Understanding the alignment of galaxy pairs which LG-like systems with the cosmic web is crucial not only for constraining the general processes of galaxy formation and evolution \citep[e.g.,][]{1992ARA&A..30..705B,2015ARA&A..53...51S,2018ARA&A..56..435W,2020MNRAS.491.4294K} but also for interpreting and placing limits on the formation history and spatial configuration of the LG. If galaxy pairs preferentially align along filaments, this may indicate that mergers and interactions are more likely to occur along these structures, or that anisotropic accretion plays a significant role in assembling systems like the LG. Furthermore, such alignments can provide valuable observational constraints on the flow of gas and dark matter within the cosmic web, and on the role of environment in shaping galaxy properties \citep[e.g.,][]{2005MNRAS.363....2K,2009Natur.457..451D}.

In this study, we aim to systematically analyze the spatial relationship between the alignment of galaxy pairs and their surrounding large-scale environment using a large sample of observational data. Our objectives include quantifying the degree of alignment between galaxy pair connections and filament orientations, investigating how this alignment depends on the physical properties of galaxy pairs and environmental density, and exploring the underlying physical mechanisms. By comparing our results with predictions from recent numerical simulations and high-resolution observations, we hope to provide new insights into the co-evolution of galaxies and large-scale structures.

This paper is organized as follows. In Section 2, we describe the observational data and filament catalogues used in our analysis， and outlines the methods employed to quantify the alignment angle between galaxy pairs and the cosmic filament. In Section 3, we present our results on the spatial alignments and their dependence on galaxy and environmental properties. Finally, Section 4 and 5 summarizes our conclusions and discusses the implications of our findings.

\begin{figure}[!htp]
    \centering
    \plotone{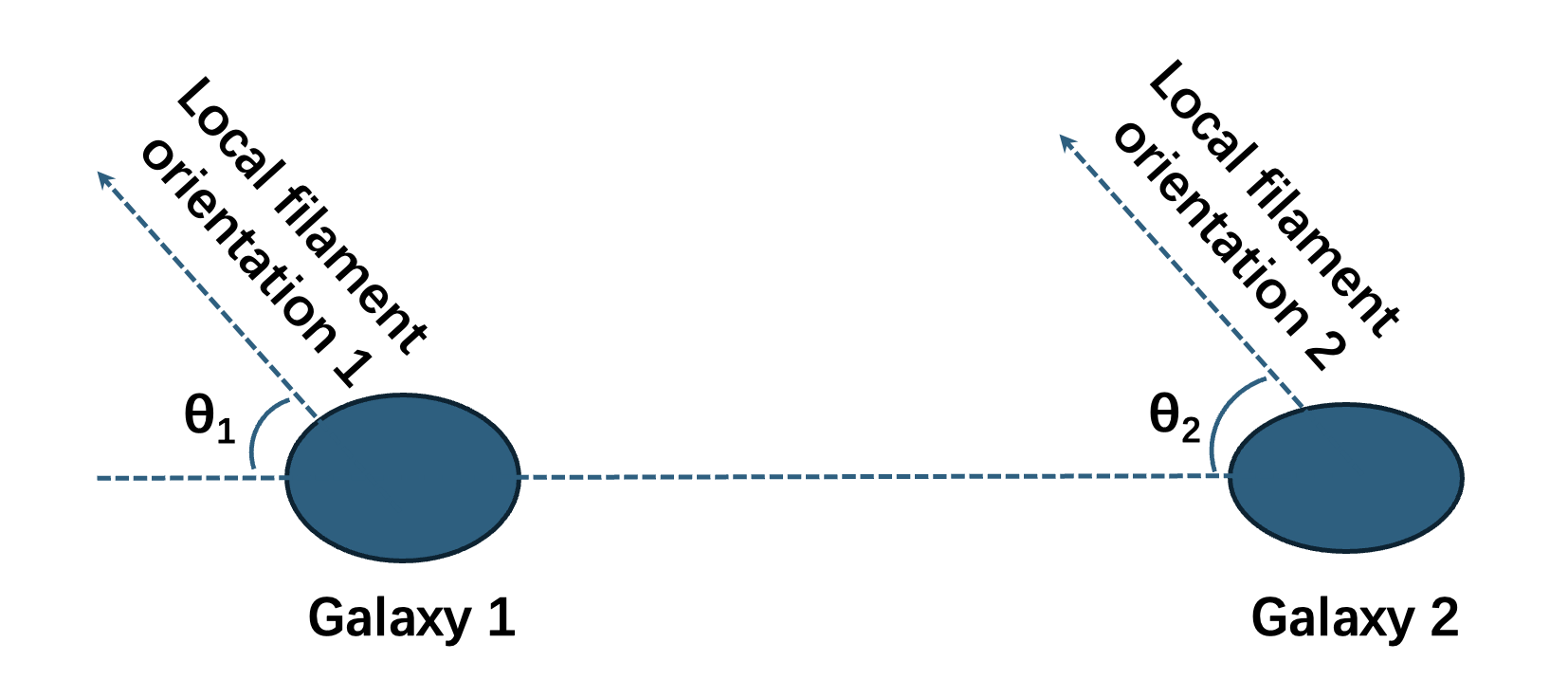}
    \caption{Illustration of the alignment angles between a galaxy pair and the local cosmic filament. The two circles represent the galaxies in the pair, connected by the pair connecting line. The arrows at each galaxy indicate the local filament orientations. The angles $\theta_1$ and $\theta_2$ denote the alignment angles between the pair connecting line and the filament orientations at each galaxy.}
    \label{fig:angle_def}
\end{figure}

\section{Data and Method} \label{sec:method}

\subsection{Observational Galaxy and Filament Data}

The galaxy sample analyzed in this study is drawn from the galaxy catalog constructed by \citet{2017A&A...602A.100T}, based on the SDSS Data Release 12 \citep{2011AJ....142...72E,2015ApJS..219...12A}. This catalog provides a comprehensive set of galaxy properties, including positions, redshifts, and group membership information. Galaxy stellar masses $M_\star$ are estimated from the $r$-band absolute magnitudes and g-r colors using mass-to-light ratio equation $\rm log(M_\star/L_r)=1.097(g-r)-0.306$ \citep{2003ApJS..149..289B}.

The large-scale filamentary structure of the cosmic web is adopted from the filament catalog used in \citet{2021NatAs...5..839W} to study the filament spin, which is also based on the SDSS galaxy distribution. Filaments are identified using the Bisous model, a marked point process algorithm developed by \citet{2014MNRAS.438.3465T,2016A&C....16...17T}. This method models the filamentary network as a collection of small, randomly oriented cylindrical segments that are iteratively connected to trace the underlying cosmic web. The resulting filament catalog provides, for each galaxy, the local filament orientation, the distance to the nearest filament axis, and the association between galaxies and filaments. 

In this work, we cross-match the galaxy catalog with the filament catalog to assign filament properties--such as orientation and distance--to each galaxy. Only galaxies with reliable filament associations are included in the analysis, enabling us to explore how galaxy properties vary with their location within the cosmic web. Our study focuses on the three-dimensional positions of galaxies, their stellar masses, colors, morphological types, and their spatial relationship to the filamentary structure, including the alignment of galaxy pairs relative to the local filament orientation. Detailed descriptions of the Bisous model implementation and the galaxy properties can be found in \citet{2014MNRAS.438.3465T, 2017A&A...602A.100T}, and the data are publicly available to facilitate reproducibility.

\subsection{Pair Selection and Angle Measurement}\label{method:def}


To investigate the possible alignment between galaxy pairs and the large-scale 
filament, we adopt a selection procedure inspired by the properties of the MW 
and M31 system.  Specifically, we select galaxy pairs whose stellar masses and three-dimensional separations are comparable to those of the MW-M31 pair. 
Our selection criteria for LG-like galaxy pairs are as follows:
\begin{itemize}
\item The stellar mass range for each galaxy in a pair is set to $0.5$--$\rm 2.0 \times 10^{12}~M_\odot$ \citep[e.g.,][and references within]{2015MNRAS.453..377W,2020SCPMA..6309801W}
\item the pair separation is required to be within $0.5$--$1.5$~Mpc. 
\end{itemize}
The narrow stellar-mass window ensures that our analysis focuses on nearly equal-mass systems (median mass ratio $q \approx 1.01$), minimizing potential biases driven by stellar mass differences within the pair.
The separation range is chosen with reference to \citet{2016ApJ...830..121L}, who adopted a maximum separation of $1.0$~Mpc when selecting Local Group-like systems from the SDSS sample, and \citet{2020MNRAS.498.2968L}, who used $1.2$~Mpc as the maximum separation for simulated Local Group analogs in the Hestia simulations. We have tested both of these maximum distance thresholds and found that they do not affect our final results. Therefore, the subsequent analysis is based on a slightly broader maximum separation of $1.5$~Mpc which ensures a sufficiently large sample size for statistical analysis. This selection yields  a fiducial sample of 3,087 galaxy pairs.

For each selected pair, we extract the three-dimensional positions \(\mathbf{x}\) \citep[see][for calculation details]{2017A&A...602A.100T} and the local filament orientations \(\mathbf{f}\) from the filament catalog. The pairwise connecting line \(\mathbf{r}\) is defined as the difference between the positions of the two galaxies. Figure~\ref{fig:angle_def} schematically illustrates the galaxy pairs, their filament orientations, and the angle definitions. Each pair consists of two members, labeled `galaxy 1' and `galaxy 2', each associated with its local filament orientation, either 1 or 2.

We calculate the angle \(\theta_i\) between the connecting line \(\mathbf{r}\) and the filament orientation \(\mathbf{f}_i\) at the position of galaxy \(i\) (\(i=1,2\)), yielding two angles per pair. Formally,  
\[
\theta_i = \arccos \left( \frac{\mathbf{r} \cdot \mathbf{f}_i}{|\mathbf{r}| \, |\mathbf{f}_i|} \right),
\]  
where \(\mathbf{r} = \mathbf{x}_2 - \mathbf{x}_1\) connects galaxy 1 to galaxy 2, and \(\mathbf{f}_i\) is the filament orientation at galaxy \(i\). Since the filament orientation is an axis without a preferred direction, \(\theta_i\) is restricted to \(0^\circ \leq \theta_i \leq 90^\circ\) by taking the smaller angle, ensuring the alignment measurement is independent of the filament orientation's sign.

To test the null hypothesis that, in the absence of intrinsic alignment, galaxy pair directions are independent of the local filament orientations, we construct a conditional randomization control sample in which the filament orientations are kept fixed while only the pair directions are randomized. The procedure is as follows:
\begin{itemize}
    \item keep the local filament orientation(s) $\mathbf{f}_i$ fixed (and the pair locations and $d_{\rm fila}$ unchanged),
    \item draw an isotropically oriented unit vector for the pair direction $\hat{\mathbf{r}}_{\rm rand}$ (uniform on the unit sphere),
    \item compute the corresponding unsigned alignment angle(s)
    \[
        \theta_{{\rm rand},i}=\arccos\!\left(\left|\hat{\mathbf{r}}_{\rm rand}\cdot \mathbf{f}_i\right|\right), \quad 0^\circ \le \theta_{{\rm rand},i}\le 90^\circ.
    \]
\end{itemize}
We repeat this procedure $N_{\rm rand}=1000$ times per pair and aggregate over all pairs to build the random-reference distributions (e.g., mean angle, CDF, and the fraction with $\theta<39.4^\circ$).

The expected mean angle for an isotropic distribution of vectors in three-dimensional space is $57.3^\circ$. For a detailed theoretical derivation of this value, we refer readers to \cite{2025ApJ...987L..30W}, which provides a comprehensive explanation. Angles smaller (larger) than $57.3^\circ$ indicate an alignment (perpendicular) trend.

We report the significance using $\sigma_{\theta}$, defined as the deviation of the observed statistic from the random ensemble in units of the random standard deviation. When using the sample mean angle $\bar{\theta}_{\rm obs}$ as the statistic, we compute
\[
\sigma_{\theta}
\equiv
\frac{\bar{\theta}_{\rm obs} - \left\langle \bar{\theta}_{\rm rand} \right\rangle}
{\sigma\!\left(\bar{\theta}_{\rm rand}\right)}
\approx
\frac{\bar{\theta}_{\rm obs} - 57.3^\circ}{\sigma\!\left(\bar{\theta}_{\rm rand}\right)} \,,
\]
where $\left\langle \bar{\theta}_{\rm rand} \right\rangle$ and $\sigma\!\left(\bar{\theta}_{\rm rand}\right)$ are estimated from the $N_{\rm rand}$ randomized catalogs. and we additionally use bootstrap resampling 1000 times of the observed pairs to report 95\% confidence bands for the empirical CDFs.

\begin{figure*}[!htp]
    \centering
    \plotone{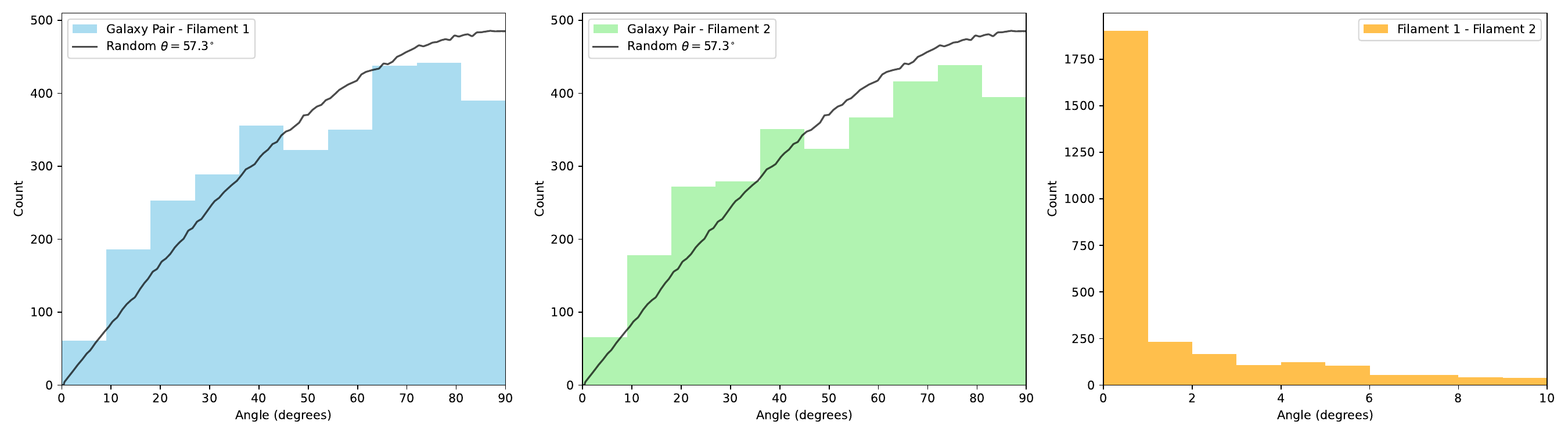}
    \caption{Distributions of alignment angles for galaxy pairs of fiducial sample. The three panels display histograms of the angle between the galaxy pair connecting line and the filament orientation at the first galaxy (left), the angle between the pair connecting line and the filament orientation at the second galaxy (middle) , and the angle between the filament orientations at the two galaxies (right). All angles are restricted to the range 0$^\circ$ to 90$^\circ$. For a random distribution of orientations in three-dimensional space, the expected mean angle is approximately 57.3$^\circ$. The black solid lines in the left and middle panels indicate the mean alignment angle distribution expected from random trials, where the pair orientations are randomized while keeping the filament orientations fixed.}
    \label{fig:check}
\end{figure*}

\subsection{Coherence of Galaxy Pairs and filament orientations}\label{sec:check}

One might notice that the local filament orientations associated with the two member galaxies in a pair can differ, potentially influencing the alignment results. As illustrated in Figure~\ref{fig:angle_def}, to ensure that the alignment outcomes are not dependent on this labeling, we conduct a detailed analysis of the angle distributions between the pair connecting line and the filament orientations at the locations of both galaxies. Figure~\ref{fig:check} presents histograms of these angles for galaxy pairs.

The left and middle panels show the distributions of the angles between the pair connecting line and the filament orientation at galaxy 1 and galaxy 2, respectively. The black solid lines in the left and middle panels indicate this random expectation, providing a baseline for comparison. The observed distributions for both galaxies deviate similarly and significantly from the random baseline, indicating a consistent alignment pattern that is independent of which galaxy in the pair is considered. 

The right panel of Figure~\ref{fig:check} shows the distribution of angles between the filament orientations at the two galaxy positions. The predominance of very small angles demonstrates that the filament orientations at the two galaxies are highly coherent, supporting the assumption that the galaxy pairs reside within the same filamentary structure. This coherence justifies using the filament orientation at either galaxy's position as a representative measure for the alignment analysis. 

Therefore, in all subsequent analyses, we treat galaxy 1 and galaxy 2 symmetrically and use the filament orientation at either galaxy's location interchangeably without affecting the results.

\section{Results} 
\label{sec:results}

\subsection{Alignment Signal of the Fiducial Sample}

\begin{figure}[!htp]
    \centering
    \plotone{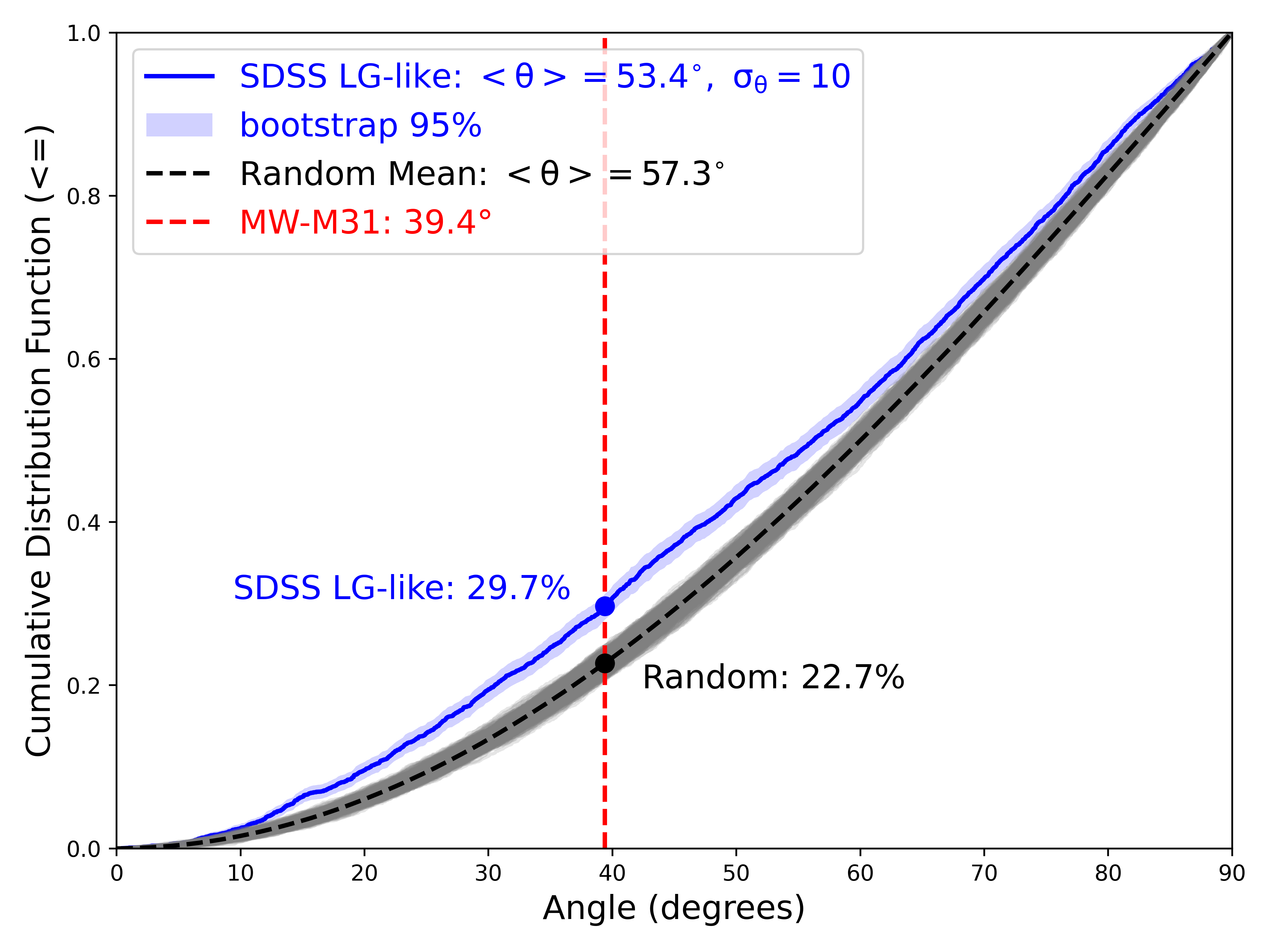}
    \caption{Cumulative distribution functions (CDFs) of the alignment angles between galaxy pair connecting lines and their corresponding local filament orientations. The blue solid line shows the observed CDF for selected 3,087 LG-like galaxy pairs in SDSS. 
The blue shaded region denotes the 95\% bootstrap confidence band by resampling the galaxy-pair sample with replacement 1000 times.
The gray lines represent CDFs from 1,000 random realizations where pair connecting line are randomized but filament orientations are kept fixed, illustrating the null hypothesis of no alignment. The black dashed line is the mean CDF of these random samples. The vertical red dashed line marks the angle of $39.4^{\circ}$, corresponding to the MW-M31 pair in the local Universe. Blue and black dots indicate the fraction of observed and random pairs with alignment angles smaller than MW-M31 case, respectively. }
    \label{fig:fig_all}
\end{figure}

We begin by examining the alignment signal between galaxy pair connecting lines and their corresponding local filament orientations in the fiducial sample. Figure~\ref{fig:fig_all} presents the cumulative distribution functions (CDFs) of the alignment angles. The observed CDF of the SDSS LG-like sample is shown as the blue solid line, representing the distribution of alignment angles measured between the vector connecting each galaxy pair and the orientation of the local cosmic filament at their positions. The mean alignment angle for the SDSS LG-like sample is $53.4^{\circ}$. 
The blue shaded region indicates the pointwise 95\% bootstrap confidence band of the observed CDF, obtained by resampling the galaxy-pair sample with replacement 1,000 times, thereby quantifying the sampling uncertainty of the observed distribution.

To assess the significance of this observed alignment signal, we compare it against a set of 1,000 randomized realizations, depicted as gray lines, in which the pair connecting lines are randomized while the filament orientations remain fixed. This procedure effectively simulates the null hypothesis of no intrinsic alignment between galaxy pairs and filaments. The black dashed line indicates the mean CDF of these random samples, serving as a baseline expectation for a purely random distribution of alignment angles. The observed mean angle deviates from the random expectation by approximately 10$\sigma$, indicating a highly significant alignment signal. Here, $\sigma_{\theta}$ denotes the deviation in units of the standard deviation of the randomized statistic. The bootstrap bands confirm that this deviation is not driven by sampling noise in the observed data.

A vertical red dashed line is drawn at an angle of $39.4^{\circ}$ \citep[][]{2015MNRAS.452.1052L}, corresponding to the alignment angle of the MW-M31 pair in the local Universe, providing a physically motivated reference point. The blue and black dots on the figure mark the fractions of observed and randomized galaxy pairs, respectively, that have alignment angles smaller than this MW-M31 benchmark. Quantitatively, 29.7\% of the galaxy pairs exhibit alignment angles less than $39.4^{\circ}$, compared to only 22.7\% in the randomized samples. \textbf{This 7\% excess with $10\sigma$ in the observed fraction represents a statistically significant deviation from randomness, indicating a clear tendency for galaxy pairs to preferentially align their connecting lines with the orientations of their local filaments.} 
The observed CDF and its bootstrap band remain well separated from the randomized baseline over a broad range of angles, confirming the robustness of the alignment signal.
Such a result agree well with \cite{2015A&A...576L...5T}, and supports the hypothesis that the cosmic filament plays a fundamental role in shaping the spatial orientation of galaxy pairs.

\begin{figure}[!htp]
    \centering
    \plotone{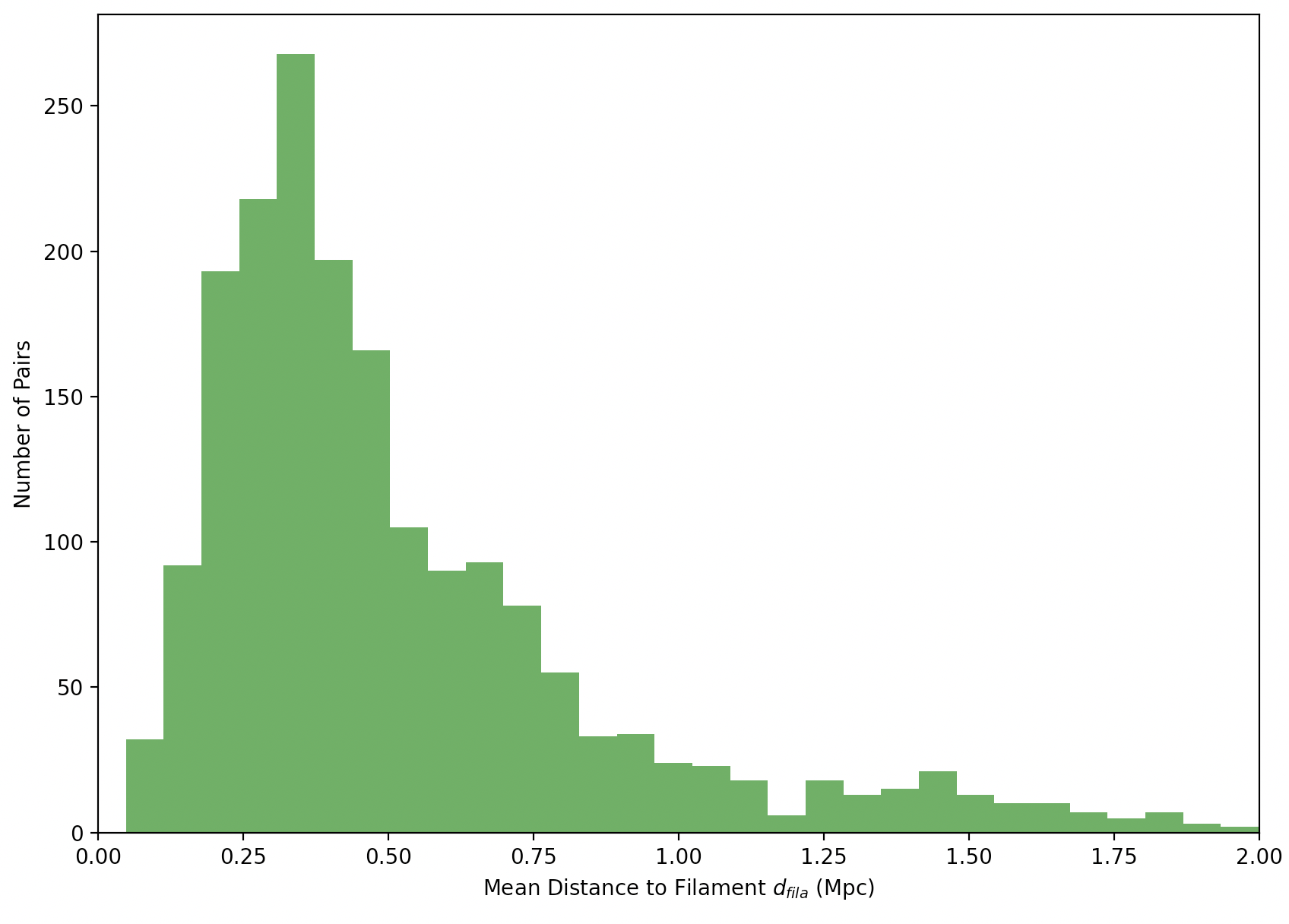}
    \caption{Histogram showing the distribution of the mean distances of galaxy pairs to their nearest filament spine. The x-axis represents the mean pair-filament distance, $d_{\rm fila}$, in megaparsecs (Mpc). The distribution peaks at approximately 0.4 Mpc and exhibits a long tail toward larger distances, indicating that most galaxy pairs reside close to filament.}
    \label{fig:pair_dist}
\end{figure}

\begin{figure}[!htp]
    \centering
    \plotone{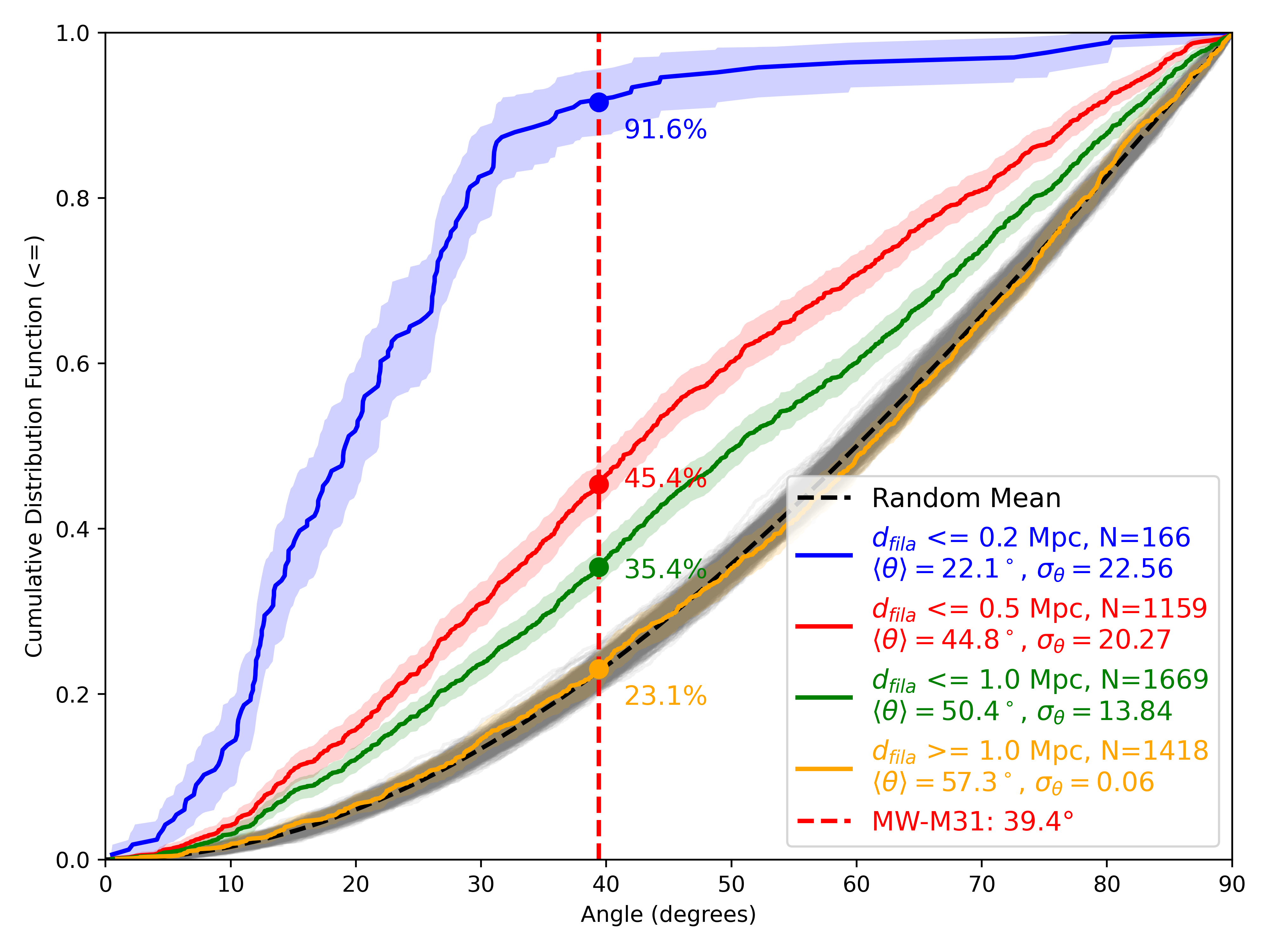}
    \caption{Similar as Fig.~\ref{fig:fig_all}, but show the effect of distance ($d_{\rm fila}$) between galaxies pairs to filament spine. The blue curve shows the CDF for pairs with $d_{\rm fila} \leq 0.2$ Mpc, the red for $d_{\rm fila} \leq 0.5$ Mpc, the green for $d_{\rm fila} \leq 1.0$ Mpc, and the orange for $d_{\rm fila} \geq 1.0$ Mpc. The colored shaded regions denote the corresponding 95\% bootstrap confidence bands for each curve, respectively. The legend lists the number of galaxy pairs in each subsample, and the confidence levels relative to randomized control samples, respectively. The 1,000 gray lines represent CDFs from random realizations, and the black solid line is the mean CDF of these random samples. The fraction of each subsample with alignment angles smaller than the MW-M31 case, namely $39.4^{\circ}$ shown in vertical red dashed line, is indicated by the corresponding points and annotated text, respectively.}
    \label{fig:fila_dist}
\end{figure}

\begin{figure}[!htp]
    \centering
    \plotone{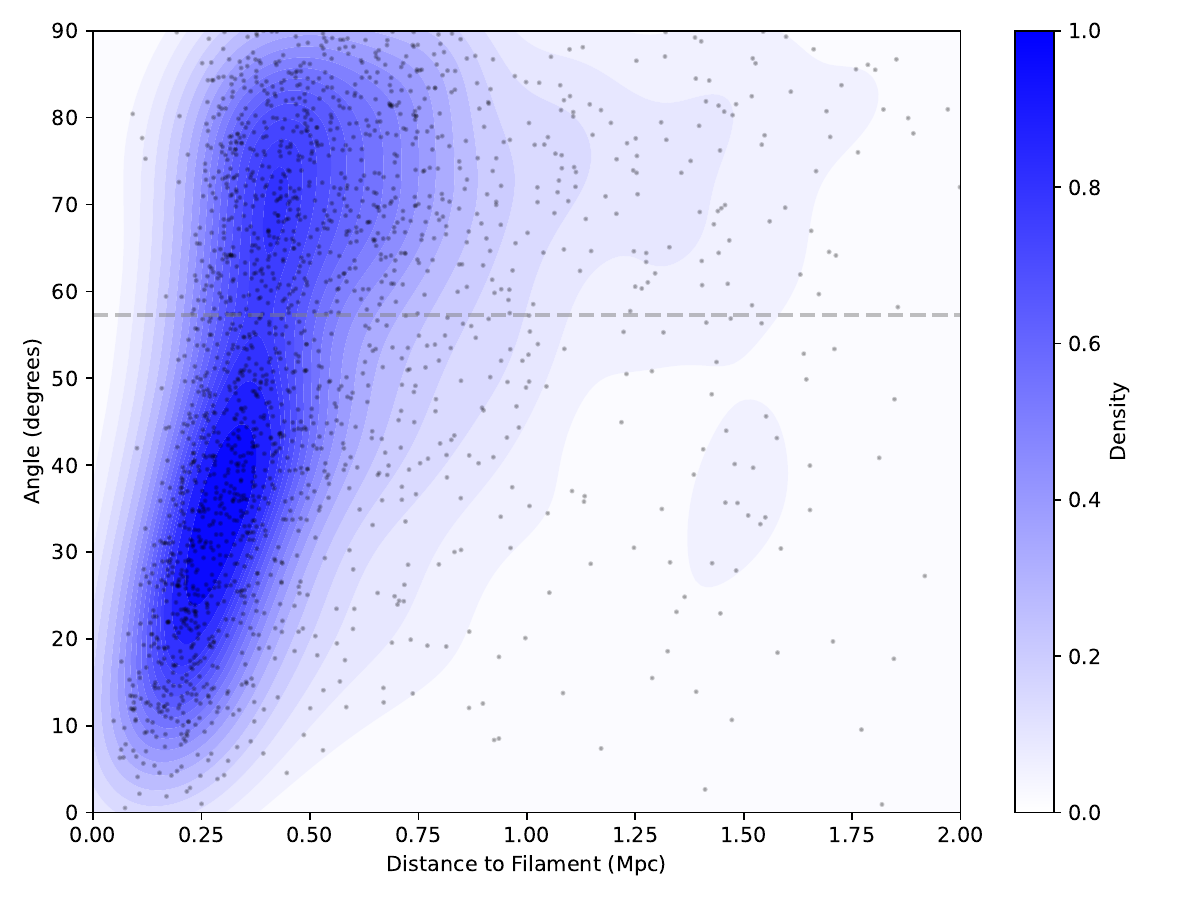}
    \caption{Scatter plots with Kernel Density Estimate (KDE)-based density maps illustrating the alignment angles versus galaxy pair - filament distances. The y-axis represents the alignment angle between the galaxy pair connecting line and the local filament orientation. The x-axis represents the distance of the galaxy pair to the filament spine.     
    Each gray dot corresponds to an individual measurement. The background color gradient and overlaid contour lines represent the number density of points, with the color bar indicating the density scale. The grey dashed horizontal line indicates the angle ($57.3^{\circ}$)  derived from the random trials.}
    \label{fig:fila_dist_count}
\end{figure}

In addition to the primary focus on the alignment between galaxy pairs and filaments, we investigated potential correlations with other galaxy properties such as stellar mass, color, and morphological type. These analyses reveal only weak or negligible dependencies of the alignment signal on these intrinsic properties. Instead, the dominant factor influencing the alignment strength is the distance of the galaxy pairs to the nearest filament spine. Motivated by this finding, the subsequent analyze concentrate on exploring how the alignment varies as a function of pair-filament distance, which emerges as the key environmental parameter driving the observed trends.

\subsection{Dependence on Filament Distance}

Building on the overall alignment results, we now turn to a more detailed investigation of how the alignment signal depends on the spatial distance ($d_{\rm fila}$) of galaxy pairs to their nearest filament spine. This focus is motivated by previous studies showing that galaxy properties such as stellar mass, star formation rate, and color exhibit strong environmental dependence related to galaxy–filament distance \citep{2017MNRAS.465.3817M,2024JCAP...02..012N}. To set the stage for the alignment analysis by distance bins, we first present the distance distribution of pairs.

Figure~\ref{fig:pair_dist} displays the histogram of the mean distances ($d_{\rm fila}$) between galaxy pairs and their associated filaments. The distribution is sharply peaked around $0.3\sim0.4$ Mpc and declines steeply beyond 0.5 Mpc, indicating that the majority of galaxy pairs we selected reside relatively close to the filamentary spine. This distribution motivates the distance ranges adopted in the subsequent CDF comparison (Figure~\ref{fig:fila_dist}) and provides context for the pair-by-distance map (Figure~\ref{fig:fila_dist_count}).

Having established the distance distribution of pairs, we next examine how alignment strength varies across those distance bins. 
Figure~\ref{fig:fila_dist} presents the cumulative distribution functions (CDFs) of alignment angles for galaxy pairs grouped by their distance to filaments.
The colored shaded regions denote the corresponding pointwise 95\% bootstrap confidence bands for each curve, respectively, obtained by resampling each subsample with replacement 1,000 times, enabling us to quantify how the strength of the alignment signal varies with filament distance.

The blue curve in Figure~\ref{fig:fila_dist} represents the CDF for galaxy pairs located within 0.2 Mpc of the filament spine. These pairs exhibit a mean alignment angle of $22.14^{\circ}$ and a highly significant confidence level of 22.56$\sigma$ relative to randomized samples. Notably, 91.6\% of these pairs have alignment angles smaller than the MW-M31  benchmark of $39.4^{\circ}$, which corresponds to an excess fraction of 68.9\% above the 22.7\% fraction expected from random orientations. This substantial excess indicates a very strong preferential alignment of galaxy pairs with their local filaments at inner side of filaments.

For galaxy pairs within 0.5 Mpc of filaments (red curve), the mean alignment angle increases to $44.81^{\circ}$, with a confidence level of 20.27$\sigma$. Here, 45.4\% of pairs are aligned more closely than the MW-M31 angle, yielding an excess fraction of 22.7\% over the random baseline. This still represents a statistically significant alignment, though weaker than that observed for the closest pairs.

Pairs within 1.0 Mpc (green curve) show a further increase in mean alignment angle to $50.37^{\circ}$, with a confidence level of 13.84$\sigma$. The fraction of pairs aligned better than MW-M31 decreases to 35.4\%, corresponding to an excess of 12.7\% relative to random. This trend suggests a gradual weakening of the alignment signal with increasing distance from the filament spine.

Finally, galaxy pairs located more than 1.0 Mpc away from filaments (orange curve) have a mean alignment angle of $57.34^{\circ}$, nearly identical to the random expectation of approximately $57.3^{\circ}$, and a negligible confidence level of 0.06$\sigma$. The fraction of pairs aligned better than MW-M31 is 23.1\%, only marginally above the random fraction by 0.4\%, indicating no significant alignment in this regime. The bootstrap confidence interval for this subsample overlaps substantially with the randomized baseline, confirming the absence of a detectable alignment signal at large distances.

Overall, these results reveal a clear and strong dependence of galaxy pair alignment on their distance to cosmic filaments. The excess fractions of aligned pairs decrease systematically with increasing distance from filaments, highlighting the critical role of the cosmic filamentary environment in shaping the spatial orientation of galaxy pairs. 
The bootstrap confidence bands remain well separated from the randomized baseline for pairs within 1.0 Mpc, while converging to the random expectation beyond 1.0 Mpc, demonstrating the robustness of the distance-dependent alignment trend.
This gradient in alignment strength supports theoretical models in which filaments act as preferred channels for galaxy formation and interaction, guiding the anisotropic accretion \citep{2014MNRAS.443.1274L,2015ApJ...813....6K,2022MNRAS.516.4576D} of galaxy systems.

To complement the cumulative analysis presented above, we further investigate how the alignment signal varies across different distances to the filament spine by examining the full distribution of alignment angles as a continuous function of pair–filament distance. While the cumulative excess fractions and confidence levels clearly show a decrease in alignment strength with increasing distance, they do not reveal how the alignment angles change progressively with distance to filament spine, nor whether there are signs of perpendicular alignments at specific distances. In other words, cumulative statistics may average over a broad range of pair–filament separations and thereby mask potential perpendicular-orientation signals.

Guided by the cumulative alignment results (Figure~\ref{fig:fila_dist}), which indicate that alignments become statistically indistinguishable from random beyond $\sim1.0$ Mpc—with mean angles converging toward the expected isotropic value of $57.3^\circ$—we therefore focus on the $0$–$1$ Mpc regime while still examining the full range for completeness. This motivates the pair-by-distance map in Figure~\ref{fig:fila_dist_count}, which allows us to assess whether and where perpendicular alignments (i.e., angles exceeding $57.3^\circ$) emerge as a function of radius.

Figure~\ref{fig:fila_dist_count} presents this refined view by directly plotting the alignment angle of each galaxy pair against its distance to the nearest filament spine, limited to within 2 Mpc. Each data point represents an individual pair, and the underlying color map shows the local density based on Kernel Density Estimation (KDE), highlighting areas of concentrated behavior.

A clear global trend is visible: alignment angles increase sharply within 0.5 Mpc from the filament spine, transitioning from tightly clustered, low-angle values to a broader, more dispersed distribution. Beyond approximately 0.5 Mpc, the angle distribution flattens and centers around $\sim57.3^\circ$, consistent with the expectation for randomly oriented orientations in two dimensions. This behavior confirms that the filamentary influence on galaxy pair alignment is strongest within 0.5 Mpc from the filament spine and rapidly weakens at larger distances.

Importantly, the KDE map reveals a bimodal distribution in the alignment angle--distance space. First, a low-angle concentration appears near 0.3 Mpc, with alignment angles between 30$^\circ$ and 40$^\circ$, indicating a population of galaxy pairs well aligned with the filament orientation, likely tracing the coherent velocity and tidal fields within the filament core. Second, a high-angle concentration is found around 0.4--0.5 Mpc, where many galaxy pairs exhibit alignment angles in the 70$^\circ$ to 80$^\circ$ range, suggesting orientations nearly perpendicular to the filament . These pairs may correspond to galaxies infalling into filaments from surrounding sheets \citep{2017MNRAS.468L.123W} or voids \citep{2017A&A...600L...6K}, or dynamically interacting at filament boundaries.

While these two populations are distinguishable, it is worth noting that the separation between them is not sharply defined. The transition occurs around 0.4 Mpc, but the boundary is diffuse, and the overall distribution is continuous rather than discrete. This likely reflects the complex interplay of multiple physical processes—including anisotropic accretion \citep[e.g.,][]{2018MNRAS.473.1562W}, filament thickness \citep{2014MNRAS.441.2923C,2020A&A...641A.173G,2022A&A...661A.115G,2024A&A...684A..63G,2024MNRAS.532.4604W}, and projection effects—that shape galaxy orientations in a gradual rather than binary fashion.

\section{Summary and Discussion}\label{sec:s}

In this study, we have investigated the spatial alignment between galaxy pairs and their surrounding cosmic filaments by using a large observational sample derived from the SDSS galaxy and filament catalogs which based on galaxy distribution. Focusing on galaxy pairs with stellar masses and separations similar to those of the MW-M31 system, we measured the angle between the vector connecting each galaxy within a pair and the local filament orientation. Our key findings are summarized as follows:

\begin{itemize}
    \item We detect a statistically significant alignment signal between the connecting lines of galaxy pairs and the orientations of their local cosmic filaments. The mean alignment angle of the observed sample is $53.4^\circ$, which is significantly smaller than the random expectation of $\sim 57.3^\circ$, with a $10\sigma$ significance level.
    \item About 29.7\% of pairs have alignment angles smaller than the MW-M31 benchmark of alignment angle of $39.4^\circ$ to local filament, representing a relative excess of approximately 7.0\% compared to the 22.7\% expected from random.
    \item The alignment strength strongly depends on the distance of galaxy pairs to the filament spine. Pairs within 0.2 Mpc of filaments show the strongest alignment, with a mean angle of $22.1^\circ$ and 91.6\% of pairs aligned better than the MW-M31 angle, representing a $22.6\sigma$ significance. This alignment weakens with increasing distance, becoming consistent with random (mean angle $\sim 57.3^\circ$) beyond 1 Mpc.
    \item The distribution of alignment angles versus filament distance reveals a bimodal pattern near the filament core: one population of pairs is well aligned with the filament axis (angles around $30^\circ\sim40^\circ$), while another shows nearly perpendicular orientations (angles around $70^\circ\sim80^\circ$), possibly reflecting different dynamical states such as infall from sheets or voids.
\end{itemize}

These results provide compelling observational evidence that the cosmic web environment plays a fundamental role in shaping the spatial orientations and dynamical evolution of galaxy pairs. The statistically significant preferential alignment of galaxy pair connecting lines with their local filament orientations lends strong support to theoretical models within the $\Lambda$CDM framework, where anisotropic accretion \citep[e.g.,][]{2007MNRAS.381...41H,2014MNRAS.443.1274L,2015ApJ...813....6K,2022MNRAS.516.4576D} and tidal torques along filaments \citep[e.g.,][]{2016IAUS..308..437C,2018MNRAS.481.4753C,2015MNRAS.452.3369C} guide galaxy formation, angular momentum acquisition \citep[e.g.,][]{2007ApJ...655L...5A,2021NatAs...5..283M,2019MNRAS.485.5244L,2021MNRAS.502.5528L}, and interactions along filaments \citep[e.g.,][]{2018MNRAS.473.1562W}.

However, \citet{2025JCAP...01..023S} reported no significant alignment signal in the EAGLE simulation \citep{2015MNRAS.446..521S}, highlighting a potentially important discrepancy between observational findings and numerical models. This discrepancy raises critical questions regarding whether it arises from limitations in simulation resolution, baryonic physics implementations, or selection effects in observational samples, and seems to underscore the need for a deeper understanding of how large-scale structure influences galaxy pair alignments across both theory and observation.

While the large $\sigma_{\theta}$ values reported in this study indicate highly significant departures from randomness, their interpretation should emphasize the primary driver: the numerator (the difference between the observed and randomized mean alignment angles). As defined in Section~\ref{method:def}, $\sigma_{\theta}$ is computed by normalizing this difference by the standard deviation of the randomized means. In our analysis, the denominator---the standard deviation of the randomized means---is relatively stable across distance bins. Therefore, large $\sigma_{\theta}$ values predominantly arise from a substantial shift of the observed mean alignment angle away from the randomized expectation. This is particularly evident for subsamples near filament spines, where the mean angle is markedly smaller than the isotropic expectation of $57.3^{\circ}$. We note that, while a small denominator can in principle amplify $\sigma_{\theta}$, this is not the dominant effect in our case. The nonparametric bootstrap confidence bands and intervals presented in Figures~3 and~5 provide a complementary measure of uncertainty, confirming that these large numerator-driven deviations remain statistically significant beyond sampling uncertainties.

Beyond statistical significance, we further tested the robustness of our alignment 
results against possible systematic effects. A potential concern when using group catalogs to study galaxy alignments in 
filaments is that a significant fraction of galaxies may reside near nodes 
(massive groups or clusters), where the local environment differs substantially 
from the filamentary regime. To test this, we examined the dependence of the 
alignment signal on group properties and found negligible correlations with both 
group mass ($r = -0.062$) and distance from the group center ($r = -0.002$), 
indicating that node contamination does not significantly bias our results.

Another possible source of uncertainty arises from redshift-space distortions (RSD) 
in galaxy positions. Although the group catalog of \citet{2017A&A...602A.100T} 
applies a statistical correction to mitigate RSD effects, those corrections are 
not intended for resolving internal group structures along the line of sight. 
Our study, however, focuses on galaxy–filament alignments on large scales, where 
such corrected coordinates are appropriate. RSD primarily affects line-of-sight 
distances, while our alignment measurement relies on angular orientations that are 
less sensitive to these uncertainties. Robustness tests—varying the pair-separation 
thresholds, checking independence from group properties, and employing an 
alternative randomization scheme—produce consistent alignment signals. Moreover, 
bootstrap uncertainty estimates (Figures~3 and~5) further confirm the significance 
of our results. We therefore conclude that our findings are robust against 
RSD-induced uncertainties, though pair orientations with strong line-of-sight 
components may be uncertain at the $\sim10^\circ$–$20^\circ$ level.

Our analysis further reveals that the alignment strength depends sensitively on the distance to the filament spine, emphasizing the role of filaments as dynamically coherent structures that most effectively influence galaxy orientations within their core regions. The gradual weakening of alignment beyond approximately 1 Mpc suggests that the gravitational and tidal influence of filaments diminishes with distance, broadly consistent with the notion that the physical boundaries of filaments extend to around 1 Mpc \citep{2024MNRAS.532.4604W}.

A particularly intriguing finding is the bimodal distribution of alignment angles near the filament core. One population of galaxy pairs aligns closely with the filament axis, likely tracing coherent velocity flows and tidal stretching along the filament. In contrast, a second population exhibits nearly perpendicular orientations, which may correspond to galaxies infalling from surrounding sheets or voids, or dynamically interacting at filament boundaries. This complexity suggests the need for more detailed kinematic studies to disentangle these scenarios and better understand the underlying physical processes.

While our current analysis focuses on alignment angles constrained to the range of $0^{\circ}-90^{\circ}$, extending this to $0^{\circ}-180^{\circ}$ in future work—particularly by incorporating velocity vectors and relative motions—will enable a more comprehensive understanding of galaxy pair dynamics within filaments.
Such studies could reveal whether galaxy pairs preferentially fall toward or recede from filament axes, offering deeper insights into anisotropic accretion and merger histories. For example, \citet{2025ApJ...987L..30W} demonstrated that by using two vectors derived from observational data, it is possible to distinguish between parallel and anti-parallel configurations of galaxy spin relative to filament spin, shedding light on the directional coherence of accretion processes.

The weak correlations observed between alignment and intrinsic galaxy properties such as stellar mass, color, or morphology suggest that the cosmic web's influence on pair orientation is primarily geometric and dynamical rather than strongly dependent on galaxy characteristics. This somehow emphasizes the dominant role of environment-driven processes in paired galaxy evolution.

More importantly, our findings offer valuable context for understanding the formation and evolution of the Local Group, particularly the MW-M31 system. The observed preferential alignment of galaxy pairs with filaments supports scenarios in which the MW and M31 formed and evolved within a coherent filamentary environment, where anisotropic accretion and tidal forces shaped their spatial configuration and merger history. This alignment may help explain the observed orbital properties and satellite distributions around these galaxies, as well as the timing and nature of their future interaction. By placing the MW-M31 pair within the broader statistical framework of filament-aligned galaxy pairs, our results provide a benchmark for testing cosmological simulations and models of Local Group assembly.

\begin{acknowledgments}
The initial idea of this paper was conceived during the CLUES 2025 meeting in Potsdam and was further developed and implemented on the ICE train from Potsdam to Frankfurt. PW gratefully acknowledges the anonymous referee for insightful comments and constructive suggestions. 
PW acknowledge the financial support from the NSFC (No.12473009), and also sponsored by Shanghai Rising-Star Program (No.24QA2711100). This work is supported by the China Manned Space Program with grant no. CMS-CSST-2025-A03. 
\end{acknowledgments}

\bibliography{main}{}
\bibliographystyle{aasjournal}



 \end{CJK*}
\end{document}